\documentclass[12pt]{book}

\usepackage[dvips]{graphicx,color}
\usepackage{makeidx,physics,cosmology}

\makeauthorindex
\makeindex

\BookTitle{Frontier in Astroparticle Physics and Cosmology}
\CopyRight{\copyright 2004 by Universal Academy Press, Inc.}

\begin{document}

\def\sun{\odot}
\def\msun{$M_\odot$}
\def\mdot{$\dot M$}
\def\msy{$M_\odot$ yr$^{-1}$}
\def\kms{km s$^{-1}$}
\def\e#1{$\times$ $10^{#1}$ }
\def\ee#1{$10^{#1}$ }
\def\Ha{H$\alpha$}
\def\co{{$^{56}$Co}}
\def\fe{{$^{56}$Fe}\/}
\def\gsim {>\kern-1.2em\lower1.1ex\hbox{$\sim$}~}   
\def\lsim {<\kern-1.2em\lower1.1ex\hbox{$\sim$}~}   

\BookTitle{\itshape Frontier in Astroparticle Physics and Cosmology}
\CopyRight{\copyright 2004 by Universal Academy Press, Inc.}
\pagenumbering{arabic}

\chapter{
Circumstellar Interaction of Type Ia Supernova SN 2002ic}

\author{%
Ken'ichi NOMOTO, Tomoharu SUZUKI, Jinsong DENG, Tatsuhiro UENISHI \\
{\it Department of Astronomy \& Research Center for the Early
Universe, School of Science, University of Tokyo, Tokyo 113-0033,
Japan} \\
Izumi HACHISU \\
{\it Institute of Earth Science and Astronomy, University of Tokyo, Meguro--ku, Tokyo, Japan} \\
Paolo MAZZALI \\
{\it Osservatorio Astronomico di Trieste, via G.B.\& Tiepolo 11, I-34131 Trieste, Italy}
}

\AuthorContents{K.\ Nomoto, T.\ Suzuki, J.\ Deng, T.\ Uenishi, I.\ Hachisu, P.\ Mazzali} 

\AuthorIndex{Nomoto}{K.}

\section*{Abstract}

     SN 2002ic is a unique supernova which shows the typical spectral
features of Type Ia supernovae (SNe Ia) near maximum light, but also
apparent hydrogen features that have been absent in SNe Ia.  We have
calculated hydrodynamical models for the interaction between the SN Ia
ejecta and the H-rich circumstellar medium (CSM) to reproduce the
observed features of SN 2002ic.  Based on our modeling, we suggest
that CSM is aspherical (or highly clumpy) and contains $\sim$ 4-5
\msun.  Possible progenitor systems of SN 2002ic are discussed.

\section{Type Ia Supernovae and Circumstellar Medium}

     Type Ia supernovae (SNe Ia) are characterized by the lack of
hydrogen and the prominent Si line in their spectra near maximum light
and widely believed to be thermonuclear explosions of mass-accreting
white dwarfs in binary systems.  SNe Ia have been used as a ``standard
candle'' to determine cosmological parameters thanks to their
relatively uniform light curves and spectral evolution.  SNe Ia are
also the major sources of Fe in the galactic and cosmic chemical
evolution.  Despite such importance, the immediate progenitor binary
systems have not been clearly identified yet (e.g.,
\cite{lund03, nom00}).

     For a model of SN Ia progenitors, Hachisu et
al. \cite{hkn99,hknu99} proposed a single degenerate model in which
the white dwarf blows a massive and fast wind (up to $10^{-4}
M_\odot$~yr$^{-1}$ and 2000~km~s$^{-1}$) and avoids a formation of
common envelope when the mass transfer rate from the normal companion
exceeds a critical rate of $\sim 1 \times 10^{-6}~M_\sun$~yr$^{-1}$
\cite{nom82}.  Such an evolutionary phase is dubbed ``accretion wind
evolution'' instead of ``common envelope evolution.''  Such a binary
can keep its separation almost unchanged.  The white dwarf can
steadily accrete a part of the transferred matter and eventually reach
the Chandrasekhar mass.

     In the strong wind model, the white winds form a circumstellar
envelope around the binary systems prior to the explosion.  When the
ejecta collide with the circumstellar envelope, X-rays, radio, and
H$\alpha$ lines are expected to be emitted by shock heating.  Attempts
have been made to detect such emissions, but so far no signature of
circumstellar matter has been detected.

     The upper limit set by X-ray observations of SN 1992A is
\mdot$/v_{10} = (2-3)$ \e{-6} \msy~ \cite{sch98}.  Radio observations
of SN 1986G have provided the most stringent upper limit to the
circumstellar density as $\dot M / v_{10} = 1 \times 10^{-7} M_\odot$
yr$^{-1}$, where $v_{10}$ means $v_{10}= v/10$ km s$^{-1}$.  This is
still $10-100$ times higher than the density predicted for the white
dwarf winds, because the white dwarf wind velocity is as fast as $\sim
1000$ km s$^{-1}$.  For H$\alpha$ emissions, the upper limit of
\mdot/$v_{10} =$ 6 \e{-6} \msy~ has been obtained for SN 1994D.

\section {SN 2002ic}

     SN~2002ic was discovered on 2002 November 13~UT at magnitude 18.5
by the Nearby Supernova Factory search \cite{woo02}.  Hamuy et
al. \cite{ham03} reported strong Fe III features and a Si II
$\lambda$6355 line in the early-time spectra of SN~2002ic and
classified it as a SN Ia.

    However, strong H$\alpha$ emission was also observed.  The
emission was broad (FWHM $> 1000$ km~s$^{-1}$) suggesting that it was
intrinsic not to an H II region of the host galaxy but to the
supernova. The detection of H$\alpha$ is unprecedented in a SN Ia
(e.g., \cite{bra95,lund03}).

    Hamuy et al. \cite{ham03} suggested that it arose from the
interaction between the SN ejecta and a dense, H-rich circumstellar
medium (CSM), as in Type IIn SNe (SNe IIn).  If this interpretation is
correct, SN~2002ic may be the first SN Ia to show direct evidence of
the circumstellar (CS) gas ejected by the progenitor system,
presenting us with a unique opportunity to explore the CSM around a SN
Ia and the nature of the progenitor system.

\subsection {Spectroscopic Features of SN 2002ic}

\begin{figure}[ht]
  \begin{center}
    \includegraphics[height=16pc]{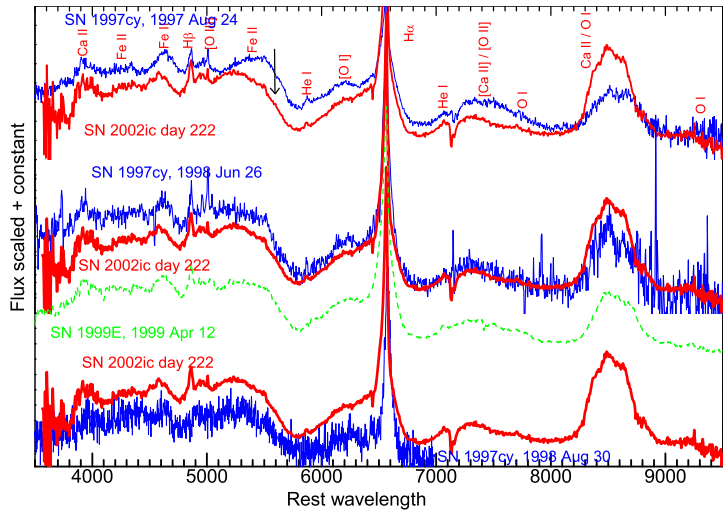}
  \end{center}
 \caption{Spectral comparison between SNe~2002ic
(red thick lines; Subaru, $\sim 222$ d), 1997cy (blue thin lines)
 and 1999E (green dashed line).
\label{fig1}}
\end{figure}

     The late-time spectrum of SN~2002ic is strikingly similar to
those of Type IIn SNe~1997cy \cite{tur98} and 1999E \cite{rig03} as
shown in Figure 1.  Spectroscopically, SN~2002ic evolve with time
significantly, in particular in the \Ha\/ line and its complex
profile. Hamuy et al. \cite{ham03} detected in the early-time spectra
an unresolved narrow \Ha\/ emission on top of a $\sim 2000$ km/s base,
which were superimposed on dominant SN Ia line features. One year
after the explosion, however, the \Ha\/ line became much more
prominent and consisted of a narrow core and a $\sim 5000$ km/s
component. Other strong features identified in Figure 1 include Ca and
O lines as broad as $\geq 10,000$ km/s and broad permitted Fe II
multiplets \cite{deng04}.

     SNe~1997cy and 1999E were initially classified as Type IIn
because they showed H$\alpha$ emission. SN~2002ic would also have been
so classified, had it not been discovered at an early epoch.
SN~1997cy ($z=0.063$) is among the most luminous SNe discovered so far
($M_{V}<-20.1$ about maximum light), and SN~1999E is also bright
($M_{V}<-19.5$).  Both SNe~1997cy and 1999E have been suspected to be
spatially and temporally related to a GRB \cite{ger00,rig03}.
However, both the classification and the associations with a GRB must
now be seen as highly questionable in view of the fact that their
replica, SN~2002ic, appears to have been a genuine SN Ia at an earlier
phase.

\subsection{Observed Light Curve of SN 2002ic}

     The {\sl UVOIR} bolometric light curve of SN 2002ic has been
constructed by Deng et al. \cite{deng04} from the available BVRI
photometry and the spectrophotometry \cite{ham03,wan04} as shown
in Figure 2.  To construct the light curve of SN~2002ic,
we first integrated the Subaru spectrum. This yielded $L = (5.9\pm
0.6)\times 10^{42}$ ergs~s$^{-1}$, corresponding to $M_{bol}\sim
-18.2$, assuming a distance of 307 Mpc ($H_0$ = 65
km~s$^{-1}$~Mpc$^{-1}$). The bolometric corrections thus estimated
was used to convert the early-time photometry in Hamuy et al.
\cite{ham03} and the late-time MAGNUM telescope photometry into
bolometric luminosities.

     The light curve of SN 2002ic is brighter at maximum and declines
much more slowly than typical SNe Ia \cite{ham03}.  The late time
light curve of most SNe is powered by the radioactive decay of
\co\/ to \fe.  The decline of SN 2002ic is much slower than the Co
decay rate, which indicates the presence of another source of
energy.

     In fact the overall light curve of SN~2002ic resembles
SNe~1999E \cite{rig03} and 1997cy \cite{tur98} (see Figure
2). We use $UBVRI$ bolometric light curves of SNe~1997cy
and 1999E for comparison \cite{tur98,rig03}. Assuming
$E(B-V)=0.06$ for the Galactic extinction (NED), SN 2002ic is only
a factor of 1.3 dimmer than SN 1997cy, but more than 100 times
brighter at late phases than typical SNe Ia. The light curve of
SN~1997cy has been modeled in the context of circumstellar
interaction \cite{tur98}, which is very likely the same energy
source for SN~2002ic.

\begin{figure}[ht]
  \begin{center}
    \includegraphics[height=16pc]{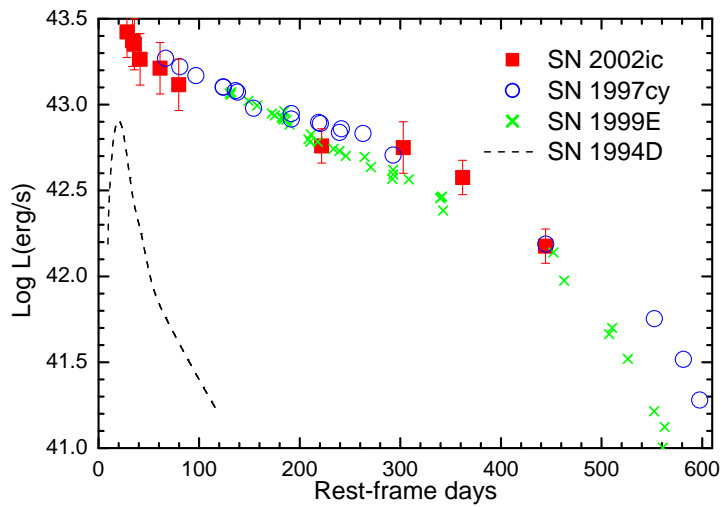}
  \end{center}
 \caption{Comparison of the $UBVRI$ bolometric light curves of
SN~2002ic (red filled squares) with those of SNe~1997cy (blue open
circles) and 1999E (green crosses), and the normal SN Ia 1994D
(black dashed line).
 \label{fig2}}
\end{figure}

\begin{figure}[ht]
  \begin{center}
    \includegraphics[height=14pc]{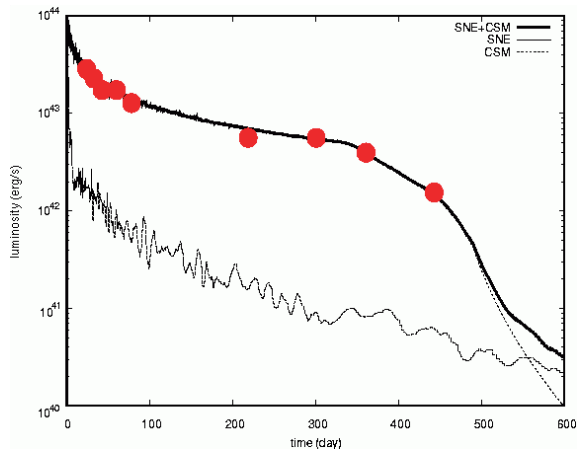}
  \end{center}
 \caption{Model light curve (black thick line) compared with
 the observation of SN 2002ic (red filled circles \cite{deng04}).
 \label{lumi}}
\end{figure}

\begin{figure}[ht]
  \begin{center}
    \includegraphics[height=14pc]{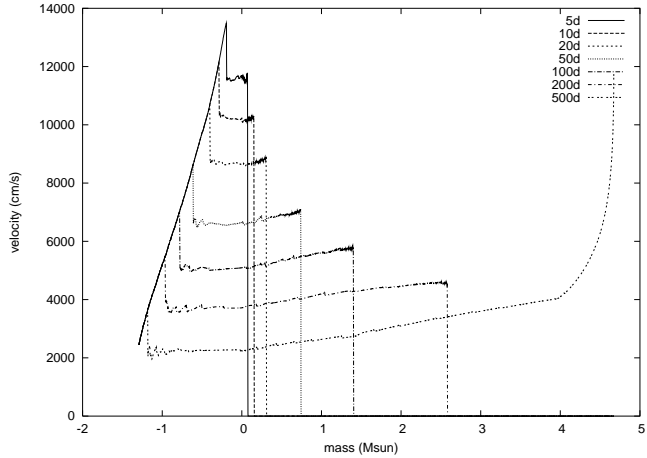}
  \end{center}
 \caption{Velocity profiles in the interacting ejecta and CSM.
 \label{velo}}
\end{figure}

\section{Circumstellar Interaction Models}

     We calculated the interaction between the expanding ejecta and
CSM (details will be seen in Suzuki et al., in preparation).  For the
supernova ejecta, we used the the carbon deflagration model W7
\cite{nom84}; its kinetic energy is $E =$ 1.3 \e{51} erg.  For CSM we
assumed the power-law density distribution:

\begin{equation}
\rho = \rho_0(r/R_0)^{-n}  \mathrm{g \ cm^{-3}}
\end{equation}
where the parameters are the radius ($R_0$) and density ($\rho_0$) of
the point where the ejecta and CSM start interacting, and the index
($n$) of the density distribution of CSM.  These parameters are
constrained from comparison with the observed light curve.  The
spherical Lagrange hydrodynamical code and input physics are the same
as in Suzuki \& Nomoto \cite{suz95}.

     When the expanding ejecta interacts with CSM, the interaction
creates the forward shock which is propagating through the CSM and the
reverse shock which is propagating through the ejecta.

     Shocked matter is heated to $T \sim$ \ee7 K for the reverse shock
and $T \sim$ \ee9 K for the forward shock. Both shocked regions
emit thermal X-rays. For the reverse shock, because of relatively
high densities in the ejecta, cooling time scale is shorter than
shock propagation so that the shocked ejecta soon forms a dense
cool shell \cite{suz95}. This dense cool shell absorbs the X-ray
and re-emits in UV-optical. This re-emitted photons are observed.
We assume that a half of the X-rays emitted in the reverse-shocked
ejecta is lost into the supernova center, and that the other half
is transferred outwardly through the cooling shell.  We also
assume that a half of the X-rays emitted in the CSM is transferred
inwardly to be absorbed by the cooling shell.  We take into
account the change in time of the column density of the cooling
shell to evaluate the X-rays absorbed by the shell and the optical
luminosity.

     Figure 3 shows the successful model for the light curve
of SN 2002ic.  Here $R_0 =$ 2 \e{14} cm, $\rho_0 =$ 4 \e{-13} g
cm$^{-3}$, and $n = 1.8$ for inner CSM of 4 \msun.  In the early
phase, the model with $n = 2.0$ (steady mass loss) declines too fast
to be compatible with the observation.  This implies that CSM around
the SN was created by unsteady mass loss of the progenitor system.

     After day $\sim$ 350, the light curve starts declining.  To
reproduce the declining part of the light curve, we add the outer CSM
of 0.7 \msun~ where the density declines sharply as $n = 6$.  This
implies that the total mass of CSM is $\sim$ 4.7 \msun.

     We note in Figure 4 that the velocity of the ejecta
decelerated by the CSM-interaction is $\lsim$ 4000 km s$^{-1}$ and too
low for the value observed in the broad spectral features ($\sim$
10,000 km s$^{-1}$).  On the other hand, in order to produce high
enough luminosity to explain the light curve, such a strong
interaction between the ejecta and CSM should occur.

     To reproduce both the light curve and the observed velocity of SN
2002ic, CSM needs to be aspherical.  Suppose the CSM is aspherical
consisting of a dense region and a thin region.  The expanding
ejecta interacting strongly with the dense region can produce high
enough luminosity to explain the light curve.  On the other hand,
the ejecta interacting with the thin region can expand still fast
enough to be consistent with the observed velocities (see also
Deng et al. \cite{deng04}). A pre-existing clumpy disk was also
suggested by Wang et al. \cite{wan04}, based on
spectropolarimetry.

\section{Discussion}

There are two possible progenitor scenarios for SN 2002ic.  One is the
the explosion of the C+O core of the massive AGB star (SN I+1/2),
where the wind from the AGB star formed the CSM.  The other is the
explosion of the white dwarf in a close binary blowing wind to create
the dense CSM (e.g., \cite{liv03,chu03}).

\subsection {Type I+1/2 Supernovae in AGB Stars}

Single star scenario is the explosion of the massive AGB star whose
C+O core becomes very close to the Chandrasekhar mass.  Before
explosion, mass loss (super-wind) from the star creates a dense CSM.
The C+O core explodes, which is called Type I+1/2 supernova, and
interacts with CSM.

To make this scenario possible, the metallicity of the system
should be low because low mass loss rate is necessary for the C+O
to grow to reach the Chandrasekhar mass before the envelope is
completely lost. Under the solar metallicity, SN I+1/2 have never
been observed. Therefore, we can explain the rarity of SN
2002ic-like event assuming that only narrow mass range of AGB
stars can explode as SNe in low metal environment.

Aspherical CSM is not unexpected for stars approaching the end of
the AGB.

\subsection {White Dwarf Winds}

Binary star scenario is the explosion of the accreting C+O white dwarf
(same as typical SNe Ia). However, the companion star is massive and
the white dwarf blows large amount of accreting gas as accretion wind
to create the dense CSM.  In this scenario, the rarity is can be
attributed to the fewness of the companion star massive enough to
produce the quite massive CSM.

As a progenitor of SN 2002ic, we need a CSM of $\sim 4.7
~M_\odot$. Such a massive CSM is possible only when the donor is as
massive as $6 - 7 ~M_\odot$.  For the model consisting of a white
dwarf and a main-sequence companion \cite{hknu99}, the mass transfer
rate from such a massive main-sequence companion reaches $\sim 1
\times 10^{-4} M_\sun$~yr$^{-1}$. Then the white dwarf blows a wind of
$\sim 1 \times 10^{-4} M_\sun$~yr$^{-1}$ and the mass stripping rate
becomes several times larger than the white dwarf wind mass loss rate
\cite{hac03kc}.

For the symbiotic model consisting of a white dwarf and a red giant or
AGB star, the wind mass loss rate can also reach $\sim 1 \times
10^{-4} M_\sun$~yr$^{-1}$. In symbiotic stars, the mass capture
efficiency by the white dwarf is observationally estimated to be as
small as one or a few percent. Therefore, only when a large part of
the red giant wind or AGB super-wind is captured by the white dwarf,
the white dwarf can blow a very massive wind of $\sim 1 \times 10^{-5}
M_\sun$~yr$^{-1}$ or more.  Then, the mass stripping rate from the red
giant or AGB star also reaches several times $10^{-4}
M_\sun$~yr$^{-1}$ or more.

Examples of the accretion wind evolution are identified as transient
supersoft X-ray sources, i.e., the LMC supersoft X-ray source
RX~J0513.9$-$6951 and its Galactic counterpart V~Sge
\cite{hac03kc}. Especially in V~Sge, a very massive wind of $\sim 1
\times 10^{-5} M_\odot$~yr$^{-1}$ has been observationally suggested
by the detection of radio.  Furthermore, the white dwarf wind collides
with the companion and strips heavily off its surface matter.  This
stripping rate reaches a few or several times the wind mass loss rate
of the white dwarf, i.e., $\sim 1 \times 10^{-4} M_\odot$~yr$^{-1}$ or
more \cite{hac03kc}. The matter stripped off has a much lower
velocity than the white dwarf wind itself and forms an excretion disk
around the binary.  Thus the model predict the coexistence of a fast
white dwarf wind blowing mainly in the pole direction and a massive
disk or a torus around the binary.  Deng et al. \cite{deng04} propose
a new classification, Type IIa supernovae, for these events.



\end{document}